\def\MET {/\!\!\!\!E_T}

\def\simlt {\,^{<}\!\!\!\!{_{\sim}}\,}
\def\ET/ {\,^{E_T}\!\!\!\!{_{/}}\,}

\documentclass[12pt]{article}
\usepackage{epsfig}

\usepackage{graphicx}
\usepackage{dcolumn}
\usepackage{bm}

\begin{document}



\centerline{
\bf 
Search for Long-lived Charged Massive
Particles in
$\bar p p$ Collisions at $\sqrt{s} = 1.8~{\rm TeV}$
}

\date{\today}

\font\eightit=cmti8
\def\r#1{\ignorespaces $^{#1}$}
\hfilneg
\begin{sloppypar}
\noindent
D.~Acosta,\r {14} T.~Affolder,\r {25} H.~Akimoto,\r {51}
M.~G.~Albrow,\r {13} D.~Ambrose,\r {37}   
D.~Amidei,\r {28} K.~Anikeev,\r {27} J.~Antos,\r 1 
G.~Apollinari,\r {13} T.~Arisawa,\r {51}
% ...
\end{sloppypar}
\vskip .026in
\begin{center}
(CDF Collaboration)
\end{center}

\vskip .026in
\begin{center}
\r 1  {\eightit Institute of Physics, Academia Sinica, Taipei, Taiwan 11529, 
Republic of China} \\
\r 2  {\eightit Argonne National Laboratory, Argonne, Illinois 60439} \\
\r 3  {\eightit Istituto Nazionale di Fisica Nucleare, University of Bologna,
I-40127 Bologna, Italy} \\
\r 4  {\eightit Brandeis University, Waltham, Massachusetts 02254} \\
\r 5  {\eightit University of California at Davis, Davis, California  95616} \\
\r 6  {\eightit University of California at Los Angeles, Los 
Angeles, California  90024} \\ 
\r 7  {\eightit University of California at Santa Barbara, Santa Barbara, California 
93106} \\ 
\r 8 {\eightit Instituto de Fisica de Cantabria, CSIC-University of Cantabria, 
39005 Santander, Spain} \\
\r 9  {\eightit Carnegie Mellon University, Pittsburgh, PA  15218} \\
\r {10} {\eightit Enrico Fermi Institute, University of Chicago, Chicago, 
Illinois 60637} \\
\r {11}  {\eightit Joint Institute for Nuclear Research, RU-141980 Dubna, Russia}
\\
\r {12} {\eightit Duke University, Durham, North Carolina  27708} \\
\r {13} {\eightit Fermi National Accelerator Laboratory, Batavia, Illinois 
60510} \\
\r {14} {\eightit University of Florida, Gainesville, Florida  32611} \\
\r {15} {\eightit Laboratori Nazionali di Frascati, Istituto Nazionale di Fisica
               Nucleare, I-00044 Frascati, Italy} \\
\r {16} {\eightit University of Geneva, CH-1211 Geneva 4, Switzerland} \\
\r {17} {\eightit Glasgow University, Glasgow G12 8QQ, United Kingdom}\\
\r {18} {\eightit Harvard University, Cambridge, Massachusetts 02138} \\
\r {19} {\eightit Hiroshima University, Higashi-Hiroshima 724, Japan} \\
\r {20} {\eightit University of Illinois, Urbana, Illinois 61801} \\
\r {21} {\eightit The Johns Hopkins University, Baltimore, Maryland 21218} \\
\r {22} {\eightit Institut f\"{u}r Experimentelle Kernphysik, 
Universit\"{a}t Karlsruhe, 76128 Karlsruhe, Germany} \\
\r {23} {\eightit High Energy Accelerator Research Organization (KEK), Tsukuba, 
Ibaraki 305, Japan} \\
\r {24} {\eightit Center for High Energy Physics: Kyungpook National
University, Taegu 702-701; Seoul National University, Seoul 151-742; and
SungKyunKwan University, Suwon 440-746; Korea} \\
\r {25} {\eightit Ernest Orlando Lawrence Berkeley National Laboratory, 
Berkeley, California 94720} \\
\r {26} {\eightit University College London, London WC1E 6BT, United Kingdom} \\
\r {27} {\eightit Massachusetts Institute of Technology, Cambridge,
Massachusetts  02139} \\   
\r {28} {\eightit University of Michigan, Ann Arbor, Michigan 48109} \\
\r {29} {\eightit Michigan State University, East Lansing, Michigan  48824} \\
\r {30} {\eightit Institution for Theoretical and Experimental Physics, ITEP,
Moscow 117259, Russia} \\
\r {31} {\eightit University of New Mexico, Albuquerque, New Mexico 87131} \\
\r {32} {\eightit Northwestern University, Evanston, Illinois  60208} \\
\r {33} {\eightit The Ohio State University, Columbus, Ohio  43210} \\
\r {34} {\eightit Osaka City University, Osaka 588, Japan} \\
\r {35} {\eightit University of Oxford, Oxford OX1 3RH, United Kingdom} \\
\r {36} {\eightit Universita di Padova, Istituto Nazionale di Fisica 
          Nucleare, Sezione di Padova, I-35131 Padova, Italy} \\
\r {37} {\eightit University of Pennsylvania, Philadelphia, 
        Pennsylvania 19104} \\   
\r {38} {\eightit Istituto Nazionale di Fisica Nucleare, University and Scuola
               Normale Superiore of Pisa, I-56100 Pisa, Italy} \\
\r {39} {\eightit University of Pittsburgh, Pittsburgh, Pennsylvania 15260} \\
\r {40} {\eightit Purdue University, West Lafayette, Indiana 47907} \\
\r {41} {\eightit University of Rochester, Rochester, New York 14627} \\
\r {42} {\eightit Rockefeller University, New York, New York 10021} \\
\r {43} {\eightit Instituto Nazionale de Fisica Nucleare, Sezione di Roma,
University di Roma I, ``La Sapienza," I-00185 Roma, Italy}\\
\r {44} {\eightit Rutgers University, Piscataway, New Jersey 08855} \\
\r {45} {\eightit Texas A\&M University, College Station, Texas 77843} \\
\r {46} {\eightit Texas Tech University, Lubbock, Texas 79409} \\
\r {47} {\eightit Institute of Particle Physics, University of Toronto, Toronto
M5S 1A7, Canada} \\
\r {48} {\eightit Istituto Nazionale di Fisica Nucleare, University of Trieste/\
Udine, Italy} \\
\r {49} {\eightit University of Tsukuba, Tsukuba, Ibaraki 305, Japan} \\
\r {50} {\eightit Tufts University, Medford, Massachusetts 02155} \\
\r {51} {\eightit Waseda University, Tokyo 169, Japan} \\
\r {52} {\eightit University of Wisconsin, Madison, Wisconsin 53706} \\
\r {53} {\eightit Yale University, New Haven, Connecticut 06520} \\
\end{center}

\begin{abstract}
We report a search for production of long-lived charged massive particles in a
data sample of $90~{\rm pb}^{-1}$ of $\sqrt{s} = 1.8 {\rm TeV} p \bar p$
collisions recorded by the Collider Detector at Fermilab (CDF).
The search uses the 
muon-like penetration and anomalously high ionization energy loss signature
expected for such a particle to discriminate it from backgrounds.
The data is found to agree with background expectations, and
cross section limits of $\cal{O}$$(1) {\rm pb}$ are derived using two reference
models, a stable quark and a stable scalar lepton.
\end{abstract}

\vskip 0.2in


Many models for new physics introduce new particles which can be long-lived
either due to a new conserved quantum number (e.g., R-parity in supersymmetry)
or because the decays are suppressed by
kinematics or couplings~\cite{theories}\cite{gluino}.
If they are electrically charged, these particles can be detected directly.
The possibility of new charged particles which are {\it absolutely} stable
is constrained by cosmological considerations and by searches for
exotic particles in stable matter~\cite{cosmo}.
However, particles which are not absolutely stable but are
long-lived on an experimental scale ($100$ ns)
are constrained only by direct searches.
The most stringent limits are set by a previous search at Fermilab's Tevatron
collider~\cite{cdf_run1} and by searches at CERN's LEP2
collider~\cite{LEPII} probing masses up to about $90~{\rm GeV}/c^2$.
In this letter, we present the results of a new search for production of
long-lived charged massive particles (CHAMPs) using a data sample of
$90~{\rm pb}^{-1}$ of $\sqrt{s} = 1.8 {\rm TeV} p \bar p$
collisions recorded by the Collider Detector at Fermilab (CDF) during 1994-5.
We search for particles with anomalously high ionization energy loss rate,
$dE/dx$, which would be produced by a slow massive charged particle.

The search can be applied to several models which
fall naturally into two distinct categories;
weakly produced particles (e.g., new leptons),
and strongly produced particles (e.g., new quarks).
The lower production cross section of weakly produced particles
only yields sufficient events for masses $\simlt 100~{\rm GeV}/c^2$ where
the background is high,
while the higher cross section of strongly produced particles
allows sensitivity at higher mass where the background is low.
The search is made as model independent as possible,
but to quantify the results we use a long-lived fourth generation quark
as a reference model for a strong production search
and Drell-Yan production of a long-lived slepton from gauge-mediated
supersymmetry breaking (GMSB) scenarios 
for a weak production search.

\begin{figure}[p]
\mbox{
\centerline{\epsfxsize=6.0in\epsffile{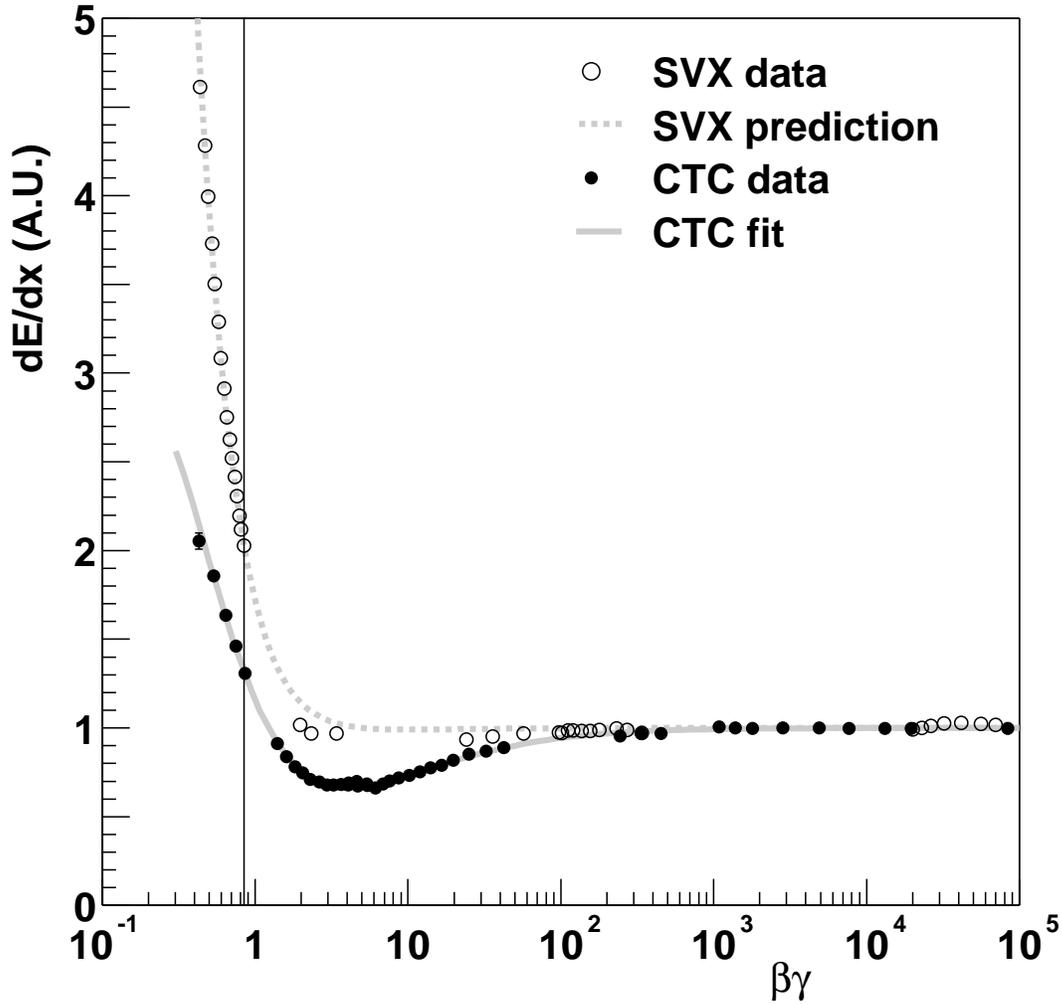}}
} \\
\caption{
$dE/dx$ measurements in control samples are compared to predictions for
SVX' (open points) and CTC (solid points).
The CTC prediction is a fit including detector effects.
The SVX' prediction is the Bethe-Bloch formula.
The agreement is good in the low and high $\beta\gamma$
regions important to this search.
}
\label{dedxfig}
\end{figure}

The CDF detector, described in detail in Ref.~\cite{CDFdescrip},
measures the trajectories (tracks) and transverse momenta~\cite{coords},
$p_T$, of charged particles in the pseudorapidity region $|\eta| < 1.1$
with the central tracking chamber (CTC) and silicon vertex detector (SVX),
which are immersed in a 1.4 T solenoidal magnetic field.
Up to 54 time-over-threshold measurements made by the CTC for each track
determine the $dE/dx$ with an average resolution of $13\%$.
The charge deposited in each of the four layers of the SVX provides a second
measure of the $dE/dx$ with an average resolution of $18\%$~\cite{dedx}. 
Control samples with well identified particle types are used to calibrate
the $dE/dx$ measurements at different velocities;
electrons and muons from $W$ boson decay at high velocity
($\beta\gamma > 100$),
muons from $J/\psi$ decay and pions $K_S$ decay at intermediate velocity,
and protons and deuterons from secondary interactions in the beampipe 
at low velocity ($\beta\gamma < 1$).
Fig.~\ref{dedxfig}
shows the comparison of these measurements to the predictions.
Electromagnetic and hadronic calorimeters, located outside the superconducting
solenoid, measure energy in segmented $\eta-\phi$ towers and
identify electron candidates.
Drift chambers for muon identification are situated outside the
$\ge 5.3$ interaction lengths ($\lambda_{\rm int}$) thick calorimeters and
behind an additional
$\ge 3.5 \lambda_{\rm int}$ thick steel absorber.

Three different trigger data sets are used for this search.
A muon trigger selects events with hits in the muon chambers 
which match a track with $p_{T}>12~{\rm GeV}/c$ in the CTC within 5$^{\circ}$.
A massive particle can penetrate the calorimeters and pass the muon trigger
even if it is strongly interacting because the energy lost in
hadronic interactions with the relatively light nucleons is too small
to initiate a shower. The CHAMP mass is $>100$ times the nucleon mass so the
energy available in the center-of-mass frame falls below the threshold for
single pion production~\cite{penet}.
Only triggers in the region $|\eta|<0.6$ are used
because at larger $|\eta|$ timing requirements that
assume $\beta=1$ are used to reduce backgrounds from beam losses.

The second trigger selects events with missing transverse energy
($E_{T}\hspace{-0.17in}/\hspace{0.1in}$) $> 35~{\rm GeV}$, which can
arise since the CHAMPs will penetrate the calorimeter without fully
depositing their energy~\cite{gluino}.
This trigger also provides acceptance for events containing neutrinos,
as is possible in GMSB models where CHAMPs are produced along with neutrinos
from the cascade decay of a heavier particle.
An electron trigger, which selects events containing electron candidates 
within the range $|\eta| < 1.1$ and with $E_T > 18~{\rm GeV}$,
provides additional acceptance for these cascade decays,
as does the muon trigger,
since charged leptons may be produced in these decays as well.

The search selects charged particle tracks with
$|\vec p| \ge 35~{\rm GeV}/c$ and $|\eta| < 1$ which have
sufficient hits in the CTC and SVX
to reduce backgrounds from misreconstructed tracks.
The $35~{\rm GeV}/c$ momentum cut is chosen because for lower momentum a CHAMP
in the mass range of interest ($M > 100~{\rm GeV}/c^2$) would be moving too
slowly to be efficiently reconstructed.
The SVX and CTC $dE/dx$ measurements are each required to be
larger than the values expected for a particle with $\beta\gamma = 0.85$.
In the region $\beta\gamma \le 0.85$,
$dE/dx \propto 1/\beta^2$ to a good approximation.
That allows calculation of a measured mass, $M_{dE/dx}$,
from the $dE/dx$ and the momentum.
The mass resolution is measured to be $20\%$ using a calibration sample of
protons and deuterons.
The search is performed for different assumed mass, M, between 
$100$ and $270~{\rm GeV}/c^2$ with $10~{\rm GeV}/c^2$ steps.
At each step, $M_{dE/dx}$ is required to be $> 0.6 \times M$, a $2 \sigma$
cut.
When combined with the $dE/dx$ cut,
this provides additional background rejection at lower momentum.
To be considered in the weak production search,
tracks must additionally pass an isolation cut requiring less than
$4~{\rm GeV}$ of calorimeter energy or total track $p_T$ within a cone of
$\sqrt{|\Delta\eta|^{2}+|\Delta\phi|^{2}} = 0.4$ around the track.

\begin{figure}[p]
\mbox{
\centerline{\epsfxsize=6.0in\epsffile{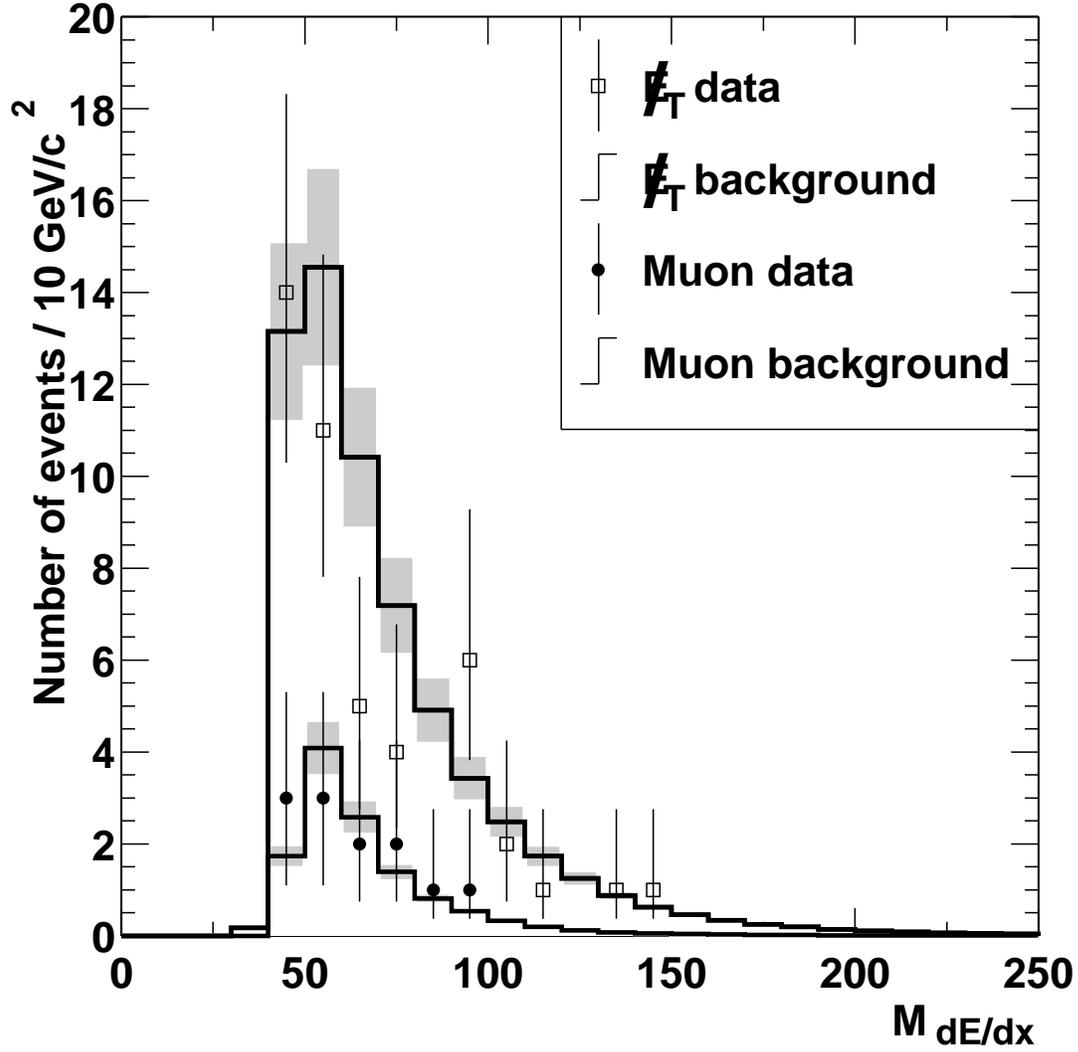}}
} \\
\caption{
Observed M$_{dE/dx}$ distribution for tracks passing all the cuts
for the strong production search in the muon trigger and $\MET$ trigger data
samples~\cite{poisson}. 
The curves are the expected background distributions which have an
uncertainty of about $15\%$ which is shown by the gray bands.
}
\label{Mass_dist}
\end{figure}

Backgrounds arise from tracks for which the $dE/dx$ measurement
fluctuated high or included extra ionization from an unreconstructed
overlapping particle.
To determine the background, we use a control sample which is identical to
the search sample but at lower momentum
($20 < |\vec{p}| < 35~{\rm GeV}/c$) where signal would not contribute.
The fake rate, defined as the fraction of tracks in the control sample with
$dE/dx$ measurements high enough to correspond to $\beta\gamma \le 0.85$, is
measured to be $\cal{O}$$(10^{-4})$
for all the different trigger datasets described above.
The momentum dependence of the fake rate within the
control sample matches expectations which allows us to extrapolate
the fake rate to the high momentum signal region.
The probability of a high fluctuation in the $dE/dx$ distribution obtained from
this fake rate is used to scale the number of candidate tracks, which pass all
selections except the $dE/dx$ requirement, to obtain background predictions of
$12 \pm 2$ tracks in the muon trigger dataset
and $63 \pm 9$ in the $\MET$ trigger dataset.
The expected mass distribution for fake tracks in the signal region is shown
in Fig.~\ref{Mass_dist}.
It is obtained by folding the momenta of the tracks into the $dE/dx$
distribution from the control sample with the assumption that large values
of $dE/dx$ are due to high mass particles.
In the data, $12$ and $45$ tracks pass all cuts for the muon and $\MET$
trigger datasets respectively.
Their mass distribution, also in Fig.~\ref{Mass_dist},
shows no significant excess over the predicted background.

For the weak production search, the isolation cut reduces the background to
$0.85 \pm 0.25$, $4.0 \pm 2.8$, and $0.7 \pm 0.5$ tracks in the muon,
$\MET$, and electron trigger datasets respectively.
In the data, $0$, $1$, and $0$ tracks are observed in these samples.

The signal efficiencies are determined using Monte Carlo simulation programs
and control data samples. 
The muon trigger efficiency is $80.5 \pm 3.0\%$,
and the track selection efficiency is $51.3 \pm 2.5 \%$,
dominated by acceptance in the SVX.
The tracking efficiency decreases at low velocity, $\beta\gamma<0.4$,
due to drift time limits in the CTC track finding algorithms.
This is measured with a sample of deuterons which are produced
from secondary interactions in the beampipe.
The efficiencies of the cuts on the kinematic variables $|\eta|$, 
$|p|$, $\beta\gamma$ ($dE/dx$) and mass are model dependent.
To set generally applicable limits,
we determine these efficiencies using easily quantifiable reference models.
For the strong production
case we use
a long-lived fourth generation quark
calculated with the Pythia Monte Carlo program~\cite{pythia}.
The total efficiency increases from $1.5$ to $2.9\%$ over
the mass range $100 - 270~{\rm GeV}/c^{2}$ for a charge ${\frac{2}{3}}e$ quark ($U$)
and from $0.8$ to $1.6\%$ for a charge $-{\frac{1}{3}}e$ quark ($D$).
The charge asymmetry of the efficiency arises from 
the light quark (u, d, s) contributions to the fragmentation;
$U \bar s$ and $U \bar d$ mesons are charged while only the $D \bar u$ meson is charged.
Furthermore, although a massive quark would efficiently penetrate the
calorimeters, the hadron containing it can undergo charge exchange from
interactions in the calorimeter which replace the light quark in it,
and a ${\frac{2}{3}}e$ quark is more likely to
remain in a charged hadron and be detected by the muon chambers.
The efficiency for this 
depends on the s quark suppression which is taken to be
$30\%$~\cite{ssuppression}.
The uncertainty from this effect,
estimated by taking half of the efficiency difference obtained if 
every hadron is assumed to interact,
is $20\%$ for $q={\frac{1}{3}}e$ and 
$13\%$ for $q={\frac{2}{3}}e$.
Other systematic uncertainties are
$4\%$ for trigger efficiency, $5\%$ for track selection, $4\%$ for luminosity,
and $7\%$ from the choice of CTEQ3M~\cite{CTEQ3M}
as the parton distribution function.
The total systematic uncertainties on efficiency are
$23\%$ and $17\%$ for $q={\frac{1}{3}}e$ and $q={\frac{2}{3}}e$ respectively.

\begin{figure}[p]
\mbox{
\centerline{\epsfxsize=6.0in\epsffile{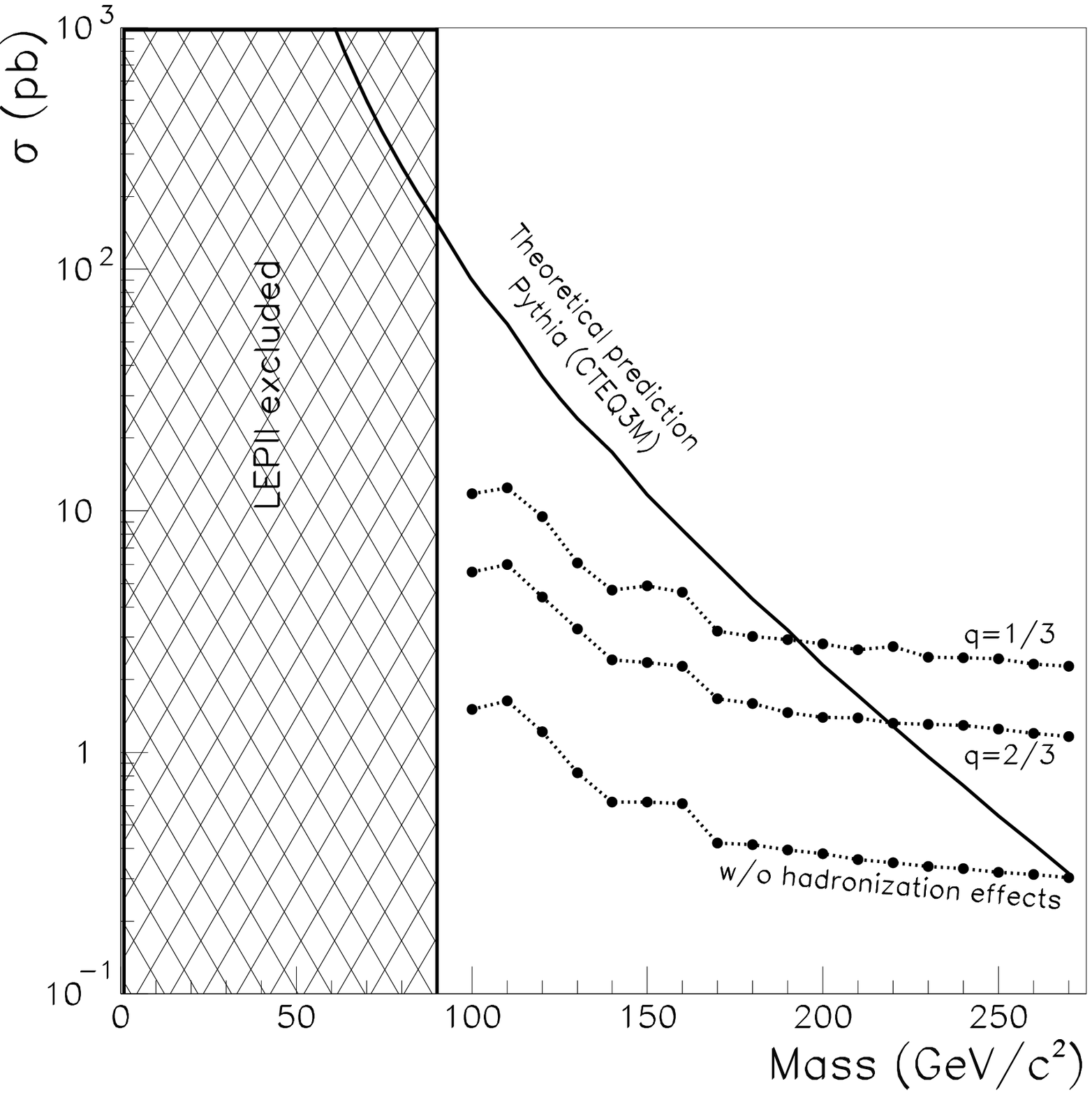}}
} \\
\caption{
Limits set at $95\%$ confidence level on the production cross section
of long-lived fourth generation quarks are compared
to the theoretical prediction.}
\label{strong_limits}
\end{figure}

Figure~\ref{strong_limits} shows the cross-section limits we derive
as a function of mass.
From comparison with the expected cross-section,
we derive mass limits at $95 \%$ confidence level of $M > 190~{\rm GeV}/c^2$ for 
$q={\frac{1}{3}}e$ and $M > 220~{\rm GeV}/c^2$ for $q={\frac{2}{3}}e$.
The charge exchange effects described above could be different for other models.
To ease comparison with other models,
we include in Fig.~\ref{strong_limits} a limit calculated
without these effects.
These limits are based on data collected with the muon trigger.
The $\MET$ trigger dataset is also searched since it could provide
sensitivity to signal, but the $\MET$ trigger efficiency depends
critically on the calorimeter's response to a CHAMP which is very uncertain.
This makes any cross-section calculations unreliable, so the $\MET$ trigger
dataset is not included in the limit calculation for the strong production
search.

For the weak production search, the muon trigger and track quality cut
efficiencies are similar to the strongly interacting case.
The efficiencies of the model dependent kinematic cuts are estimated using
as a reference model the Drell-Yan pair-production of stable sleptons
calculated with the SPythia Monte Carlo program~\cite{spythia}.
The total efficiency varies from $2.5 \%$ to $4.5 \%$ over the mass
range $80 - 120~{\rm GeV}/c^{2}$.
The systematic uncertainties on these efficiencies are similar to the
strongly interacting case, without the charge exchange uncertainty.
The cross-section limits obtained for direct slepton production
range from $1.3~{\rm pb}$
at $M=80~{\rm GeV}/c^{2}$ to $0.75~{\rm pb}$ at $120~{\rm GeV}/c^2$ .
The expected cross section is over an order of magnitude below
this level of sensitivity.
Stable sleptons can also be produced from cascade decays of heavier
particles such as charginos.
Such decays would also produce charged leptons and neutrinos,
and the electron and $\MET$ trigger data samples are searched to add
sensitivity to these decays.
The efficiency for this is very model dependent,
and we quantify it only for a single point in the GMSB parameter space 
which makes the three charged sleptons nearly degenerate with
masses $\sim 105~{\rm GeV}/c^2$, slightly above the
existing limits~\cite{GMSBpoint}.
The modified kinematics from the decays increases the
efficiency to $6.7\%$ for the muon trigger data set.
Including the $\MET$ and electron triggers increases it to 
$8.2\%$.
The $\MET$ trigger and isolation requirement introduce additional systematic
uncertainties from the modelling of initial and final state radiation,
making the total systematic uncertainty 12.5$\%$.
When cascade decays from all production modes are included,
the cross-section limit is lowered to $550 {\rm fb}$ compared to the
model prediction of $80 {\rm fb}$.

In summary, we have searched for long-lived charged massive particles in
$90~{\rm pb}^{-1}$ of data at CDF.
No excess over background was observed.
We derive cross-section limits using reference models for
the two cases of strongly and weakly produced particles.
In the strongly interacting case,
these limits extend the excluded mass region to about $200~{\rm GeV}/c^2$.

We thank the Fermilab staff and the technical staffs of the participating
institutions for their vital contributions.
We thank S.~Ambrosanio, J.L.~Feng, J.F.~Gunion, and P.Q.~Hung for useful
theoretical discussions.
This work is supported by the
U.S. Department of Energy and the National Science Foundation;
the Natural Sciences and Engineering Research Council of Canada;
the Istituto Nazionale di Fisica Nucleare of Italy;
the Ministry of Education, Science and Culture of Japan;
the National Science Council of the Republic of China;
and the A.P. Sloan Foundation.


\begin{thebibliography}{99}
\bibitem{dedx}
 FERMILAB-CONF-97/430-E.  A low-end truncated mean is used for both the CTC and SVX to suppress Landau fluctuations.
\bibitem{GMSBpoint} The GMSB parameters for the point used are
 $N_5=3$, $M=64~TeV$, $\Lambda=32~TeV$, $tan(\beta)=3$, $\mu>0$.
\end{thebibliography}

\end{document}